\documentstyle[emulateapj]{article}
\begin{document}
\title{Dust vs. Cosmic Acceleration}
\author{Anthony N. Aguirre}
\affil{Harvard-Smithsonian Center for Astrophysics, 60 Garden 
Street, Cambridge, MA 02138; aaguirre@cfa.harvard.edu}
\submitted{Accepted by The Astrophysical Journal Letters} 
 
\begin{abstract}
        
  Two groups have recently discovered a statistically significant
  deviation in the fluxes of high-redshift type Ia supernovae from the
  predictions of a Friedmann model with zero cosmological constant.
  In this Letter, I argue that bright, dusty, starburst galaxies would
  preferentially eject a dust component with a shallower opacity curve
  (hence less reddening) and a higher opacity/mass than the observed
  galactic dust which is left behind.  Such dust could cause the
  falloff in flux at high-$z$ without violating constraints on
  reddening or metallicity.  The specific model presented is of
  needle-like dust, which is expected from the theory of crystal
  growth and has been detected in samples of interstellar dust.
  Carbon needles with conservative properties can supply the necessary
  opacity, and would very likely be ejected from galaxies as required.
  The model is not subject to the arguments given in the literature
  against grey dust, but may be constrained by future data from
  supernova searches done at higher redshift, in clusters, or over a
  larger frequency range.

\noindent {\em Subject headings:} cosmology: observations -- dust, extinction --
radiative transfer

\end{abstract}

\section{Introduction}

Recently, two separate groups have interpreted their observations of
type Ia supernovae as evidence for acceleration in the cosmic
expansion, presumably caused by a nonzero cosmological constant (Riess
et al. 1998, henceforth `R98', and Perlmutter et al. 1998).  Using
the supernovae as standard candles (after correction for the relation
between luminosity and light-curve shape), both groups find a
progressive dimming of supernovae at high redshift relative to the
predictions of a flat, matter-dominated model, or even an open model
with zero cosmological constant.  There are at least two explanations
for this dimming which are unrelated to the cosmological parameters:
evolutionary effects and dust. The two observational groups have
expended considerable effort attempting to account for such
systematics, but this letter argues that the obscuration of distant
supernovae by a cosmological distribution of dust is {\em not} ruled
out, by developing a simple intergalactic dust model which is
reasonable in its creation and dispersement, has reasonable physical
properties, and could cause the observed effect without violating
constraints such as those noted in R98.

The standard way to estimate dust extinction, used by the supernova
groups, is to apply a relation between reddening and extinction
derived from Galactic data (see, e.g., Cardelli, Clayton \& Mathis
1989).  This relation is a result of the frequency dependence of the
opacity curve of the absorbing dust, which can be measured well for
Galactic dust at wavelengths $\lambda \la 100\ \mu$m, and fit by
theoretical models of graphite and silicate dust in the form of small
spheres with a distribution of radii (Draine \& Lee 1984; Draine \&
Shapiro 1984).  Because the opacity of this `standard' dust falls off
rather quickly with increasing wavelength for $\lambda \ga 0.1\ 
\mu$m, attenuated light is significantly reddened.

The weakness of this method when applied to a new situation is the
necessary assumption that the same extinction-reddening relation
holds, even though the dust may be of a different species; standard
techniques would not correct for dust which causes extinction without
significant reddening.  For the same reason, an effectively uniform
intergalactic distribution of non-reddening dust could remain
undetected by reddening studies such as those of Wright \& Malkan
(1988) and Cheng, Gaskell \& Koratkar (1991).

\section{A Specific Model with Carbon Needles}

To make this idea concrete, let us begin with the theory of dust
formation.  Very small dust grains are believed to form in a
vapor-solid transition, via a process of crystal formation; these
grains may then coagulate into larger ones.  In small grain
formation, nuclei form first, and then they grow as surface nucleation occurs
on their faces.  Surface nucleation creates steps which grow along the
face, adding a new layer and increasing the crystal size.  But
environments of low supersaturation (as may commonly occur
astrophysically) strongly inhibit surface nucleation.  In such cases,
grains may still grow by the mechanism of a `screw dislocation' (see
e.g. Frank 1949, Sears 1955), or by growth of a rolled-up platelet (Bacon
1960), forming one-dimensional `needles' like those commonly found in
laboratory experiments (e.g. Nabarro \& Jackson 1958.)

Moreover, needles can grow rapidly where `spherical' dust cannot; thus
in some situations needles can out-compete spherical dust for
available metals.  This reasoning led Donn \& Sears (1963) to predict
that interstellar dust could be dominated by needle-type dust.  These
predictions were partially borne out by the discovery of estatite
needles in captured interstellar dust (Bradley, Brownlee \& Veblen
1983.)  The needles were not the dominant component, but their
discovery does demonstrate that vapor-solid transitions occur
astrophysically, and that astrophysical needles can form by the same
mechanism as in laboratories (they contained screw dislocations.)
Conducting needles are physically interesting because they act as
antennas, absorbing electromagnetic radiation more effectively than
standard dust.  Several authors have proposed models in which such
grains thermalize the microwave background (see, e.g., Wickramasinghe
et al. 1975; Wright 1982; Hawkins \& Wright 1988).

\begin{figure}
\plotone{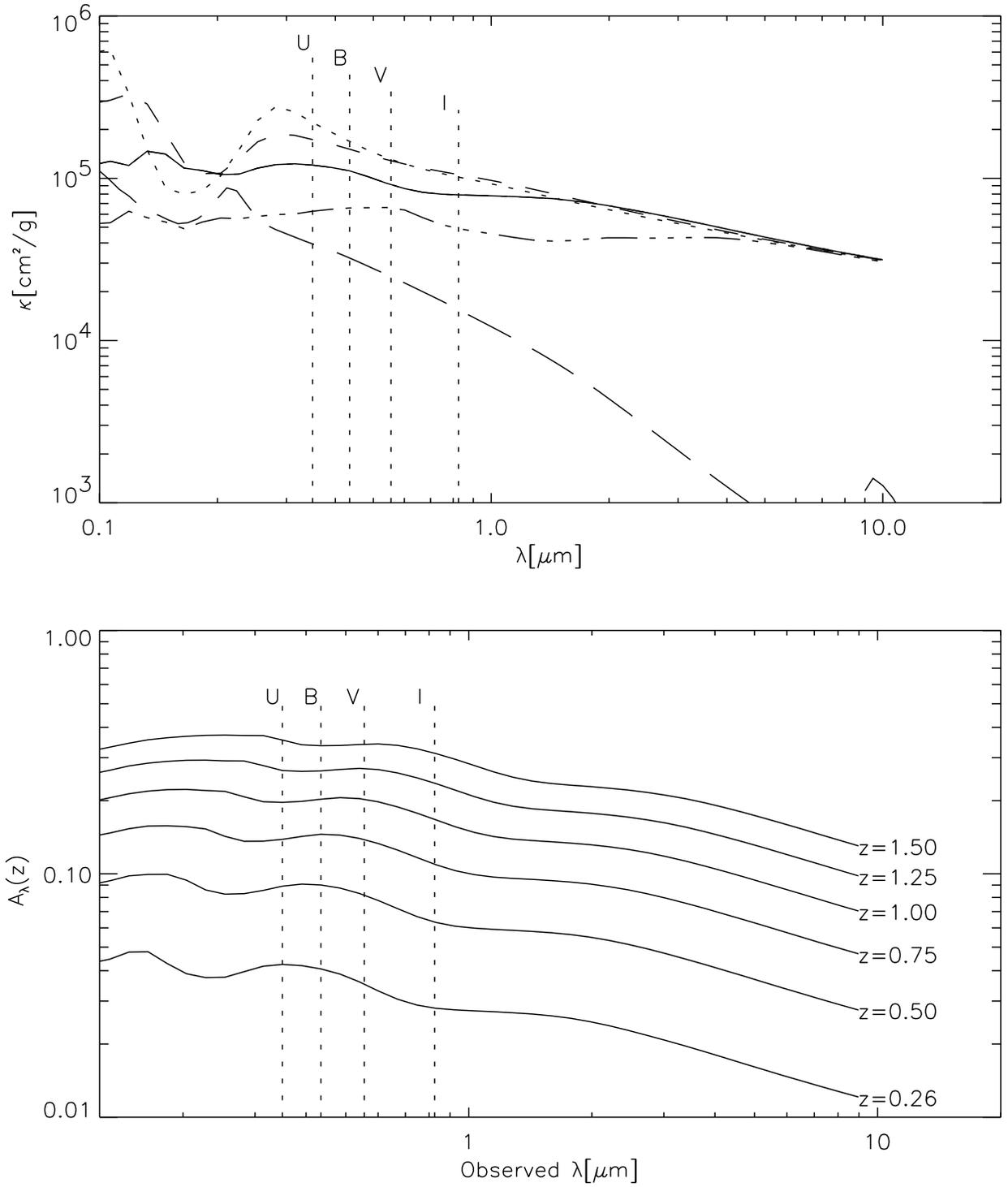}
\caption{{\bf Top:} Carbon needle opacity curves for aspect ratio distribution
  $n\propto (L/d)^{-1}$ and diameters $d=0.02, 0.04, 0.1, 0.2\ \mu$m
  (dotted, dashed, solid and triple-dot-dashed, respectively).  Also
  included are central U, B, V and I frame wavelengths and a `Draine
  \& Lee' dust curve (long-dashed) calculated using the Laor \&
  Draine (1993) method.  {\bf Bottom:} Integrated extinction to
  various redshifts for $d=0.1\ \mu$m needles for $\Omega=0.2$ and
  $\Omega_{needle} = 10^{-5}.$}
\label{fig-needopac}
\end{figure}

I have calculated the extinction cross section for needles at $\lambda
\le 10\ \mu$m using the `discrete dipole approximation' (see, e.g.,
Draine 1988) as implemented in the publicly available DDSCAT package
and using the accompanying graphite dielectric constants
$\epsilon_\perp$ and $\epsilon_\parallel$ (see Laor \& Draine 1993).
Following Wickramasinghe \& Wallis (1996), I assume that the graphite
$c$-axis is perpendicular to the needle so that $\epsilon_\perp$
applies for an electric field parallel to the needle axis (see also
Bacon 1960).  Figure \ref{fig-needopac} shows curves for various
needle diameters $d$, $0.02 \le d[\mu m] \le 0.2$, averaged over
incident radiation directions, and averaged over an aspect ratio
distribution $n(L/d) \propto (L/d)^{-1}$ with $4 \le L/d \le 32$.
This mass-equipartition distribution was chosen to represent a
shattering spectrum from longer needles; laboratory needles grow up to
$L/d \sim 1000.$ The maximal $L/d$ is somewhat arbitrary but largely
irrelevant since the short-wavelength behavior depends only weakly
on $L/d \ga 8.$ The results roughly agree with the Mie calculations
of Wickramasinghe and Wallis (1996) which use somewhat different
optical data.

Major uncertainties in the opacity model include the uncertainties in
optical data (see Draine \& Lee 1984 for discussion), an unknown
impurity content in the needles, and the unknown needle
diameter;\footnote{ The needle diameter is particularly important, but
  it is difficult justify any {\em a priori} estimate of its value.
  For the sake of the argument at hand I take $d=0.1.$ Note that a
  distribution of diameters would likely simply be dominated by the
  low-diameter cutoff.} the model given is intended to be suggestive
rather than complete.  The key point is that needles generically have
an opacity which is higher ($\kappa_V \simeq 10^5\ {\rm cm^2\ 
  g^{-1}}$) and less wavelength-dependent than that of standard dust.


Several works (Chiao \& Wickramasinghe 1972, Ferrara et al. 1990 and
1991, Barsella et al. 1989) have studied dust ejection from galaxies,
all concluding that most spiral galaxies could eject (spherical)
graphite dust.  These theoretical studies are supported by
observations of dust well above the gas scale height in galaxies
(Ferrara et al. 1991) and of vertical dust lanes and fingers
protruding from many spiral galaxies (Sofue, Wakamatsu \& Malin 1994).
The high opacity of needles extends over a large wavelength range,
hence they are strongly affected by radiation pressure and are even
more likely be ejected than spherical grains.  In a magnetized region,
charged grains spiral about magnetic field lines.  Magnetized gas may
escape from the galaxy via the Parker (1970) instability, or grains
may escape by diffusing across the lines during the fraction of the
time that they are uncharged; see, e.g., Barsella et al. (1989).  Once
free of the galaxy\footnote{A grain leaving the disk would be
  sputtered by gas in the hot galactic halo, but the resulting mass
  loss is $ <20\%$ for the 0.01 $\mu$m silicate spheres and even less
  for faster moving or larger grains (Ferrara et.  al.  1991), so the
  effect on the needles would be very small.}, needles would rapidly
accelerate and could reach distances of 1 Mpc or more.

Following Hoyle \& Wickramasinghe (1988), we estimate the time
required for needle ejection and dispersement as follows.  A grain
with length $L$, cross section $d^2$, specific gravity $\rho_m$ and
opacity $\kappa$ in a anisotropic radiation field will attain a
terminal velocity $v$ given by equating the radiative acceleration
$\kappa F/c$ to the deceleration due to viscous drag of $a \approx
(4v^2 / \pi d)(\rho_{gas} / \rho_{m}) + {\cal O}(u^2/v^2).$ Here, $F$
is the net radiative flux, and $u=(3kT_{gas}/m_H)$ and $\rho_{gas}$
are the gas thermal speed and density. Values applicable for needles
in our Galaxy are $\kappa \simeq 10^5 {\rm\ cm^2\ g^{-1}}$, $\rho_m
\simeq 2\ {\rm g\ cm^{-3}}$, $\rho_g \simeq 10^{-24}\ {\rm g\ 
  cm^{-3}}$, $T_{gas} \sim 100 $K, and $F/c \sim 10^{-13}\ {\rm ergs\ 
  cm^{-3}}$.  These give a terminal velocity of $v \simeq 4 \times
10^5 (d/0.1 \mu{\rm m})^{1/2}\ {\rm cm\ s^{-1}}$ and a timescale to
escape a 100 pc gas layer of $\sim 2.5 \times 10^7(d/0.1 \mu{\rm
  m})^{-1/2}\ {\rm yr.}$ Outside the gas layer, the needle is subject
only to radiation pressure.  For a rough estimate we assume that the
constant acceleration $\kappa F/c \sim 10^{-8}\ {\rm cm\ s^{-2}}$ acts
for the time required for the needle to travel a distance equal to the
galactic diameter.  This takes $\sim (2Rc/\kappa F)^{1/2} \simeq
8 \times 10^7\ {\rm yr}$ for a galaxy of size $R \sim 3 \times
10^{22}\ {\rm cm}$ and leaves the needle with velocity $v \sim 2.5
\times 10^7\ {\rm cm\ s^{-1}}$.  Such a velocity will carry the needle
1 Mpc (twice the mean galaxy separation at $z=1$) in $\sim 4$ Gyr.
For comparison, the time between $z=3$ (when dust might be forming)
and $z=0.5$ (when the supernovae are observed) is 5.5 Gyr for $\Omega
= 1$ and 7.3 Gyr for $\Omega = 0.2.$ These estimates suggest that
radiation pressure should be able to distribute the dust fairly
uniformly\footnote{Using the model of Hatano et al.  (1997), R98
  argues that dust confined to galaxies would cause too large a
  dispersion in supernova fluxes.  For the dust to create less
  dispersion than observed, they must merely be `uniform' enough that
  a typical line-of-sight (of length $\sim cH_0^{-1}$) passes through
  many clumps of size $\sim D$ and separation $\sim \lambda$, i.e.
  that $1/\sqrt{cH_0^{-1}D^2/\lambda^3} \la 1$ (the observed
  dispersion in R98 is 0.21 mag, similar to the necessary extinction).
  This is easily satisfied for needles traveling $\ga 50$ kpc from
  their host galaxies.} before $z \simeq 0.5$.

Dust is known to exist in large quantities (masses $\sim 0.1\%$ of the
total galaxy mass are often inferred) in bright, high-redshift
galaxies (see, e.g., Hughes 1996).  These galaxies would preferentially
eject dust with higher opacity at long wavelengths (e.g. needles, or
fractal/fluffy grains); such grains tend to have a shallower falloff
in opacity with wavelength, hence redden less than the observed
Galactic dust.  This selection affect and the estimation of dust
escape timescales suggest that if substantial intergalactic dust
exists, it should be effectively uniform, and redden less than
standard dust.

We can compute the optical depth to a given redshift due to uniform
dust of constant comoving density using
$$
\tau_\lambda(z) = \left({c\over H_0}\right)\rho_0\Omega_{needle}\int_0^zdz'\ 
{(1+z') \kappa[\lambda/(1+z')] \over (1+\Omega z')^{1/2}}.
$$
Figure \ref{fig-needopac} shows the integrated optical depth to
various redshifts for needles with $d = 0.1\ \mu m$, for $\Omega =
0.2$, $h=0.65$ and $\Omega_{needle} = 10^{-5}.$ Using this
information, we can calculate the dust mass necessary to account for
the observations if $\Omega_\Lambda = 0.$
The difference between an $\Omega =0.2,\ 
\Omega_\Lambda=0.0$ model and a model with $\Omega =0.24,\ 
\Omega_\Lambda = 0.76$ (the favored fit of R98) is about 0.2
magnitudes at $z=0.7.$ In the $d=0.1\ \mu$m needle model this requires
$\Omega_{needle} = 1.6 \times 10^{-5}.$ Matching an $\Omega = 1,\ 
\Omega_\Lambda=0$ universe requires about 0.5 magnitudes of extinction
at $z=0.7$ and $\Omega_{needle} = 4.5\times 10^{-5}.$
 
A reddening correction based on standard dust properties, like that
used in R98, would not eliminate this effect. R98 effectively
estimates extinction using rest-wavelength (after K-correction) $B-V$
color and the Galactic reddening law.  For standard dust this would be
reasonable even for a cosmological dust distribution, since the
reddening would still occur the across the redshift-corrected $B$ and
$V$ frames.  But Figure \ref{fig-needopac} shows that this does not
hold for needles: the $d=0.1\ \mu$m needle distribution only gives
$(B-V) = 0.06 A_V$ up to $z=0.7$.  The supernova group method would
$K-$correct the $B$ and $V$ magnitudes, then convert this (rest frame)
$B-V$ into an extinction based on the Galactic $(B-V) = 0.32 A_V.$ It
would therefore not be surprising for the systematic extinction to go
undetected.

Studies of redshift-dependent reddening (e.g. Wright 1981, Wright \&
Malkan 1987, Cheng et al. 1991) in far-UV (rest frame) quasar spectra
put limits on a uniform dust component, but these are most sensitive
to high redshifts, at which the needles would not yet have formed and
uniformly dispersed.  In addition, it is clear from Figure
\ref{fig-needopac} that for thick needles the flatness of the opacity
curve would lead to a very small shift in the quasar spectral index up
to $z=1.$

Another available constraint, the metallicity of Ly-$\alpha$ clouds,
is probably not relevant; because the dust formation and ejection (due
to radiation pressure) from galaxies is independent of the enrichment
mechanism of the clouds (presumably population III enrichment or gas
`blowout' from galaxies), there is no clear connection between the
mass of metal gas in the clouds and the mass of needle dust in the
IGM\footnote{Of course, if the needles were also assumed to form in the
  population III objects their density should then relate to the
  Ly-$\alpha$ metallicity.}.
 
To estimate the fraction of carbon locked in the needle dust, we would
like to know $\Omega_Z$ at $0.5 \la z \la 3.$ The current value of
$\Omega_Z$ should be bounded above by the metal fraction of large
clusters, which are the best available approximation to a closed
system that is a fair sample of the universe.  Clusters tend to have
$\sim 1/2$ solar metallicity (e.g. Mushotsky et al. 1996), and $\sim
10\%$ of their mass in gas (e.g. Bludman 1997 for a summary), giving
$\Omega_Z \la 10^{-3}.$ This compares reasonably well with an upper
bound on universal star density estimated from limits on extragalactic
starlight (from Peebles 1993) of $\Omega_* < 0.04$: if we extrapolate
the Galactic metallicity of $\sim Z_\odot$, we find $\Omega_Z \sim
Z_{\odot}\Omega_{*} \la 4 \times 10^{-4}.$ Assuming a current
$\Omega_Z \sim 4\times 10^{-4}$ and that metals are created constantly
(conservative, given the higher star formation rate at high-$z$) in
time from $z=6$ we find (for both $\Omega = 0.2$ and $\Omega = 1$)
that $\Omega_Z(z=3) \sim 4 \times 10^{-5}$ and $\Omega_Z(z=0.5) \sim 2
\times 10^{-4}$, which agrees with recent estimates by Renzini (1998).
Even such crude approximations are very vulnerable, but suggest that
the needed amount of needle mass is reasonable.

The needle model is falsifiable in several ways.  First, the needle
opacity spectrum is not perfectly flat, especially for small $d.$
Observations over a long wavelength span might reveal a
redshift-dependent systematic change in certain colors.

Next, the needles take some minimum time to form, then more time to
achieve a uniform cosmic distribution.  Thus at high enough redshift
the dispersion in supernova brightnesses discussed in R98 appears.
Moreover, at $z=1.5$ the difference
between the $\Omega = 0.2, \Omega_\Lambda =0$ model with dust and the
$\Omega = 0.24, \Omega_\Lambda = 0.76$ model sans dust is $\simeq 0.2$ mag,
which should eventually be observable.

I shall not attempt to address the question of galaxy counts here.  As
commented in R98, grey dust would exacerbate the `problem' of
unexpectedly high galaxies counts at high-$z$, but the magnitude of
such an effect would depend upon the dust density field's redshift
evolution, and a full discussion of the galaxy count data as a
constraint on the model (requiring also an understanding of galaxy
evolution) is beyond the scope of this letter.

Galactic observations probably cannot disprove the model, since
needles with the properties most different than those of Galactic dust
would be ejected with high efficiency.  Moreover, dust with
needle-like characteristics may have been detected by COBE (Wright et.
al. 1991; Reach et al. 1995; Dwek et al. 1997) as a minor `very cold'
component of Galactic dust.  Such a component is best explained by
dust with a hitherto-unknown IR emission feature, or by fluffy/fractal
or needle dust (Wright 1993) and could represent a residual needle
component with about 0.02-0.4\% of the standard dust
mass.\footnote{The needles absorb about 5$\times$ as effectively (per
  unit mass) in the optical where most Galactic radiation resides, and
  the `very cold' component emits between .1\% and 2\% of the total
  FIR dust luminosity (Reach et al. 1995).}

On the other hand, the dust cannot escape from clusters, which have
much higher mass/light ratios, so needles formed {\em after} the
formation of a cluster should remain trapped within.  Studies of
background quasar counts (Bogart \& Wagner 1973; Boyle et al. 1988;
Romani \& Maoz 1992), cooling flows (Hu 1992), and IR emission
(Stickel et al. 1998) of rich clusters indicate extinctions $A_V \sim
0.2 - 0.4$ mag and standard dust masses of $M^{dust}_{cl} \sim
10^{10}\ M_\odot.$ Denoting by $Z_{cl}, M_{cl}$, and $M^{gas}_{cl}$
the mean cluster metallicity, total mass and gas mass, we can estimate
the fraction of metals in dust $\chi_{cl}$ to be
$$
\chi_{cl} = {M_{cl} \over M^{gas}_{cl}}
{M^{dust}_{cl} \over M_{cl}}
Z_{cl}^{-1}
\simeq 10 \times 10^{-5} / 0.01 = 0.01,
$$
using $M_{cl} \sim 10^{15}\ M_{\odot}.$ Comparing this to the
$\chi_{gal} \la 1$ typical of our Galaxy would
indicate dust destruction efficiency of $\la 99\%$ in clusters.  An
earlier calculation gave $\Omega_{needle} / \Omega_Z \ga 0.1$ for
the intergalactic needles.  Assuming the calculated dust destruction,
this predicts $M^{needle}_{cl}/M^{dust}_{cl} \sim 0.1.$ The needles
are about five times as opaque in optical as standard dust, so this
gives an optical opacity ratio of $\sim 0.5$.  If these estimates are
accurate, comparison of nearby cluster supernovae to nearby
non-cluster supernovae at fixed distance should reveal a mean
systematic difference of $A_V \ga 0.03 - 0.06$ in fluxes {\em after}
correction for reddening.\footnote{Similar arguments {\em might} apply
  to elliptical galaxies, from which dust ejection is less efficient
  than from spirals.} The Mt. Stromlo Abell cluster supernova search
(Reiss et al.  1998), currently underway, should make such an
analysis possible.  Note that uncertainties in the needle opacity
relative to standard dust will not affect the cluster prediction which
(modulo the quantitative uncertainties) should hold unless clusters
destroy needles more efficiently than standard dust.

\section{Conclusions}
        
I have argued that the reduction of supernova fluxes at high redshift
could be caused by a uniform distribution of intergalactic dust. 

Both theoretical arguments and observational evidence strongly suggest
that some dust should be ejected from galaxies.  Dust with high
opacity (especially at long wavelengths where most of the luminosity
of high-redshift starburst galaxies resides) would be preferentially
ejected.  But this is exactly the sort of dust which would both redden
less than standard dust, and require less dust mass to produce the
observed effect.  This letter develops a specific model of
intergalactic dust composed of carbon needles--a theoretically
reasonable, even expected, form of carbon dust--with conservative
properties.  The supernova data can be explained by a quantity of
carbon needles which is plausible in light of rough estimates of the
universal metal abundance.

  Because the dust distribution is effectively uniform, it does not induce a
dispersion in the supernova magnitudes, and because it absorbs more
efficiently than standard dust, it does not require an unreasonable
mass.  Finally, because the dust is created and ejected by high-$z$
galaxies, it does not overly obscure {\em very} high redshift galaxies
or quasars.  Thus the key arguments given in R98 against `grey' dust
do not apply.  The dust of the proposed model should, however, provide
independent signatures of its existence; one is a systematic
difference in fluxes between cluster and non-cluster supernovae which
may be detectable in ongoing surveys.  Finally, the needle model is
only one specific model for intergalactic dust.  Other possible `dust'
types are fractal dust (e.g. Wright 1987), platelets (e.g. Donn \&
Sears 1963; Bradley et al. 1983), hollow spheres (Layzer \& Hively 1973), or
hydrogen snowflakes.

The explanation of reduced supernova flux at high redshift described
in this letter depends upon the plausible but still speculative
assumption that the intergalactic dust distribution has significant
mass, and is dominated by grains with properties exemplified by those
of carbon needles.  The probability that this is the case should be
weighed against the severity of the demand that the explanation
favored by Riess et al. and Perlmutter et al. places on a solution
of the vacuum energy (or cosmological constant) problem: the expected
value of the vacuum energy density at the end of the GUT era must be
reduced by some as yet unknown process, not to zero, but to a value
exactly one hundred orders of magnitude smaller.

\acknowledgements 

This papers has benefited significantly from the
commentary of an anonymous referee.  I thank David Layzer, George
Field, Alyssa Goodman, Bob Kirshner, Ned Wright and Saurabh Jha for
useful discussions.

\section{Note added in proof}

Two recent papers bear upon this Letter.  Kochanek et al. (1998,
astro-ph/9811111) find that dust in early-type high $z$ galaxies
reddens {\em more} than `standard' dust.  Needles may escape
ellipticals, but the lensing technique applied to clusters would be an
excellent test of the needle model.

Perlmutter et al. (1998, Ap. J., accepted) find no statistically
significant different in mean reddening between low-$z$ and high-$z$
samples; however, the $B-V \simeq 0.01$ mag that my fiducial model
predicts fits easily within their 1-$\sigma$ errors.

\newpage

\end{document}